# SafeLS: Toward Building a Lockstep NOEL-V Core

Marcel Sarraseca[†], Sergi Alcaide[†], Francisco Fuentes[†,*], Juan Carlos Rodriguez[†],
Feng Chang[†], Ilham Lasfar[†], Ramon Canal[†,‡], Francisco J. Cazorla[†], Jaume Abella[†]

[†] Barcelona Supercomputing Center (BSC), Barcelona, Spain
[‡] Universitat Politècnica de Catalunya (UPC), Barcelona, Spain
[*] Universitat Autònoma de Barcelona (UAB), Barcelona, Spain

**Abstract**

*Safety-critical systems such as those in automotive, avionics and space, require appropriate safety measures to avoid silent data corruption upon random hardware errors such as those caused by radiation and other types of electromagnetic interference. Those safety measures must be able to prevent faults from causing the so-called common cause failures (CCFs), which occur when a fault produces identical errors in redundant elements so that comparison fails to detect the errors and a failure arises. The usual solution to avoid CCFs in CPU cores is using lockstep cores, so that two cores execute the same flow of instructions, but with some time staggering so that their state is never identical and faults can only lead to different errors, which are then detectable by means of comparison. This paper extends Gaisler's RISC-V NOEL-V core with lockstep; and presents future prospects for its use and distribution.*

## Introduction

The development process for safety-critical systems makes the risk of software and systematic hardware errors residual by design. However, random hardware faults, such as those caused by radiation, are unavoidable and appropriate safety measures must be deployed in accordance with functional safety standards (e.g., ISO26262 for automotive systems). In the context of safety-critical systems, spatial redundancy is a usual solution to detect most errors.

Solutions implementing spatial redundancy are abundant, e.g., those realizing dual (DMR) [4] or triple modular redundancy (TMR) [3]. However, spatial redundancy is not enough in the case of safety-critical systems that carry high integrity requirements. Some faults may affect clock or power signals, hence propagating the fault to all redundant elements. If those elements are designed in a way that their electrical state can be identical at any point in time, they may experience identical errors, hence leading to a failure since comparison cannot detect the errors. These failures are known as Common Cause Failures (CCFs), and safety standards impose that CCFs must be prevented.

The usual way to avoid CCFs consists of using diverse redundancy, so that common faults lead to different errors, and hence, they become detectable [1]. For instance, storage usually implements error correcting codes such as Single Error Correction Double Error Detection (SECDED) codes, where the data and the corresponding code are different (e.g., different bit count and bit values), so that even if the fault corrupts both the data and the code, the chances of detecting the error are high (e.g., if up to 1 bit of the data and 1 bit of the code are affected, detection is guaranteed with SECDED).

In the case of computing cores, the usual solution uses lockstep execution: two cores are coupled in a way that only one of them is visible at user level. Yet, internally, both cores execute the same instruction flow, but with some time staggering so that their internal electrical state differs at any point in time. As a result, a fault affecting both cores cannot produce identical errors. Commercial solutions for lockstep cores can be found from different chip vendors, such as Infineon AURIX microcontrollers. However, to our knowledge, open source realizations integrated in safety-relevant SoCs are not available yet.

This paper covers this gap by developing a lockstep version of the RISC-V based NOEL-V core by Frontgrade Gaisler, and integrating it as part of the open source SELENE SoC [2]. In particular, this paper presents key design choices taken along with some necessary context. The full design will be released open-source by the time of the summit.

## SafeLS: a Lockstep NOEL-V Core

**Context**. In the context of safety-critical systems in the automotive, space, and avionics domains, among others, it is generally admissible failing to complete the execution of an instance of a task (a.k.a. a job). This is true even if at the highest integrity level, as long as the error is detected. This is, for instance, the case of the braking system of a car.

Let us assume that the software function of the braking system has to take a decision within 200ms due to safety requirements, and it is executed every 50ms. Even if we detect the error at the end of the execution, there is enough time to reexecute and ensure a timely braking if the risk of two consecutive errors can be argued to be negligible with appropriate safety argumentation.

**Design choices**. Different alternatives exist to set the sphere of replication. Some designs aim at setting the sphere of replication at the pipeline stage level, as recently done, for instance, for a RISC-V core [5]. Such a design has some pros and cons. The main advantage is that errors are detected immediately. The disadvantage is that such a design is highly intrusive with existing cores, which requires duplicated efforts for verification and validation.

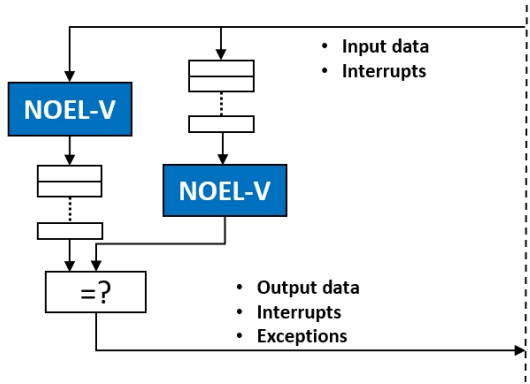

**Figure 1. SafeLS architecture**

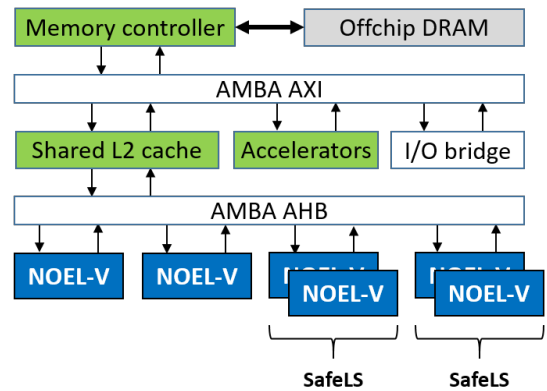

**Figure 2. SELENE SoC with 2 SafeLS and 2 regular NOEL-V cores**

Our target segment is safety-critical systems for the automotive, space, avionics, and robotics domains, among others, where, in general, errors can be managed at coarser grain than pipeline stages, as discussed before. Hence, immediate detection is not needed. As long as errors are detected when the job of a safety-critical task completes its execution, those errors do not lead to a failure, and safety requirements are preserved. Thus, we opt for setting the sphere of replication at core level, as shown in Figure 1 for our realization of the SafeLS for the NOEL-V core.

In particular, all input signals for the NOEL-V core (e.g., input data, interrupts) are replicated and forwarded to the two redundant cores. One of them (the left one in the figure) receives those inputs immediately, whereas the other receives those inputs with some programmable delay (typically 2 or 3 cycles). Then, the outcomes of the left core (e.g., output data, interrupts generated, exceptions) are delayed by the same number of cycles as the inputs of the right core, and then compared with the outcomes of the right core. If no discrepancy is detected, outcomes are delivered back to the rest of the SoC just once since, externally, only one core is visible. This is shown in Figure 2, which depicts the schematic of the SELENE SoC including 2 regular cores and 2 SafeLS components.

Upon a discrepancy, at least one of the outcomes is erroneous. If DMR is employed, as usually done in automotive systems, it is unknown which of the outcomes is correct, if any (note that a CCF could have made all outcomes be erroneous). Hence, an interrupt is raised so that the error can be properly managed at system level.

## Future Plans

As indicated before, a functional implementation of the SafeLS is already in place. This design has been verified and validated. Our plans for the near future include the release as an open-source component, both as a standalone module as well as integrated in the SELENE SoC. Our aim is to keep extending such SoC with additional features until making it be complete enough to include all features needed in a SoC for safety-critical systems with the highest integrity level in domains such as automotive, space, railway and robotics, among others.

Also, we note that first level caches of the commercial release of the NOEL-V core are properly protected against single bit upsets (SBUs). The read-only instruction cache is parity protected, so SBUs can be corrected simply by invaliding the erroneous cache line and fetching it again from upper levels in the memory hierarchy. Similarly, the data cache implements a write-through policy with the L2 cache along with parity protection, so SBUs can be detected and corrected analogously as for the instruction cache. Hence, we plan to set the sphere of replication including the cores without their first level caches to avoid the duplication of those caches. We foresee this approach to be more intrusive than our current design, but it will provide some relevant gains in terms of power and area.

## Summary

Safety-critical systems impose the use of lockstep cores to avoid CCFs. Our work delivers SafeLS, the first open-source RISC-V-based lockstep core based on the commercial Gaisler's NOEL-V and integrated into a fully-functional SoC (the SELENE SoC). The lockstep is validated and the implementation will be released open-source by the time of the summit (doing it now would violate the blindness of the submission).

This work is part of the project PCI2020-112010, funded by MCIN/AEI/10.13039/501100011033 and the European Union ``NextGenerationEU''/PRTR, and the European Union's Horizon 2020 Programme under project ECSEL Joint Undertaking (JU) under grant agreement No 877056. This work has also been partially supported by the Spanish Ministry of Science and Innovation under grant PID2019-107255GB-C21 funded by MCIN/AEI/10.13039/501100011033.